\documentclass{aastex631}

\usepackage{xcolor}
\hypersetup{
    colorlinks,
    linkcolor={red!80!black},
    citecolor={blue!80!black},
    urlcolor={magenta!80!black}
}

\newcommand\jwst{\textit{JWST}}

\usepackage{amsmath}

\received{November 22, 2023}

\submitjournal{PASP}

\shortauthors{Ma et al.}

\begin{document}

\title{\jwst's PEARLS: Improved Flux Calibration for NIRCam}

\author[0000-0003-3270-6844]{Zhiyuan Ma}
\affiliation{Department of Astronomy, University of Massachusetts, Amherst, MA 01003, USA}
\author[0000-0001-7592-7714]{Haojing Yan} 
\affiliation{Department of Physics and Astronomy, University of Missouri, Columbia, MO\,65211, USA}
\author[0000-0001-7957-6202]{Bangzheng Sun}
\affiliation{Department of Physics and Astronomy, University of Missouri, Columbia, MO\,65211, USA}

\author[0000-0003-3329-1337]{Seth H. Cohen} 
\affiliation{School of Earth and Space Exploration, Arizona State University,
Tempe, AZ 85287-1404, USA}

\author[0000-0003-1268-5230]{Rolf A. Jansen} 
\affiliation{School of Earth and Space Exploration, Arizona State University,
Tempe, AZ 85287-1404, USA}

\author[0000-0002-7265-7920]{Jake Summers} 
\affiliation{School of Earth and Space Exploration, Arizona State University,
Tempe, AZ 85287-1404, USA}

\author[0000-0001-8156-6281]{Rogier A. Windhorst}
\affiliation{School of Earth and Space Exploration, Arizona State University,
Tempe, AZ 85287-1404, USA}

\author[0000-0002-9816-1931]{Jordan C. J. D'Silva} 
\affiliation{International Centre for Radio Astronomy Research (ICRAR) and the
International Space Centre (ISC), The University of Western Australia, M468,
35 Stirling Highway, Crawley, WA 6009, Australia}
\affiliation{ARC Centre of Excellence for All Sky Astrophysics in 3 Dimensions
(ASTRO 3D), Australia}

\author[0000-0002-6610-2048]{Anton M. Koekemoer} 
\affiliation{Space Telescope Science Institute,
3700 San Martin Drive, Baltimore, MD 21218, USA}

\author[0000-0001-7410-7669]{Dan Coe} 
\affiliation{Space Telescope Science Institute, 3700 San Martin Drive, Baltimore, MD 21218, USA}
\affiliation{Association of Universities for Research in Astronomy (AURA) for the European Space Agency (ESA), STScI, Baltimore, MD 21218, USA}
\affiliation{Center for Astrophysical Sciences, Department of Physics and Astronomy, The Johns Hopkins University, 3400 N Charles St. Baltimore, MD 21218, USA}

\author[0000-0003-1949-7638]{Christopher J. Conselice} 
\affiliation{Jodrell Bank Centre for Astrophysics, Alan Turing Building,
University of Manchester, Oxford Road, Manchester M13 9PL, UK}

\author[0000-0001-9491-7327]{Simon P. Driver} 
\affiliation{International Centre for Radio Astronomy Research (ICRAR) and the
International Space Centre (ISC), The University of Western Australia, M468,
35 Stirling Highway, Crawley, WA 6009, Australia}

\author[0000-0003-1625-8009]{Brenda Frye} 
\affiliation{Department of Astronomy/Steward Observatory, University of Arizona, 933 N Cherry Ave,
Tucson, AZ, 85721-0009, USA}

\author[0000-0001-9440-8872]{Norman A. Grogin} 
\affiliation{Space Telescope Science Institute,
3700 San Martin Drive, Baltimore, MD 21218, USA}

\author[0000-0001-6434-7845]{Madeline A. Marshall} 
\affiliation{National Research Council of Canada, Herzberg Astronomy \&
Astrophysics Research Centre, 5071 West Saanich Road, Victoria, BC V9E 2E7,
Canada}
\affiliation{ARC Centre of Excellence for All Sky Astrophysics in 3 Dimensions
(ASTRO 3D), Australia}

\author[0000-0001-6342-9662]{Mario Nonino} 
\affiliation{INAF-Osservatorio Astronomico di Trieste, Via Bazzoni 2, 34124
Trieste, Italy} 

\author[0000-0002-6150-833X]{Rafael {Ortiz~III}} 
\affiliation{School of Earth and Space Exploration, Arizona State University,
Tempe, AZ 85287-1404, USA}

\author[0000-0003-3382-5941]{Nor Pirzkal} 
\affiliation{Space Telescope Science Institute,
3700 San Martin Drive, Baltimore, MD 21218, USA}

\author[0000-0003-0429-3579]{Aaron Robotham} 
\affiliation{International Centre for Radio Astronomy Research (ICRAR) and the
International Space Centre (ISC), The University of Western Australia, M468,
35 Stirling Highway, Crawley, WA 6009, Australia}

\author[0000-0003-0894-1588]{Russell E. Ryan, Jr.} 
\affiliation{Space Telescope Science Institute,
3700 San Martin Drive, Baltimore, MD 21218, USA}

\author[0000-0001-9262-9997]{Christopher N. A. Willmer} 
\affiliation{Steward Observatory, University of Arizona,
933 N Cherry Ave, Tucson, AZ, 85721-0009, USA}

\author[0000-0001-8751-3463]{Heidi B.~Hammel} 
\affiliation{Association of Universities for Research in Astronomy, 1331 Pennsylvania 
Avenue NW, Suite 1475, Washington, DC 20005, USA}

\author[0000-0001-7694-4129]{Stefanie N.~Milam} 
\affiliation{NASA Goddard Space Flight Center, Greenbelt, MD\,20771, USA}

\author[0000-0003-4875-6272]{Nathan J. Adams} 
\affiliation{Jodrell Bank Centre for Astrophysics, Alan Turing Building, 
University of Manchester, Oxford Road, Manchester M13 9PL, UK}

\author[0000-0003-0202-0534]{Cheng Cheng}
\affiliation{Chinese Academy of Sciences South America Center for Astronomy, National Astronomical Observatories, CAS, Beijing 100101, China}

\author[0000-0001-6145-5090]{Nimish P. Hathi} 
\affiliation{Space Telescope Science Institute,
3700 San Martin Drive, Baltimore, MD 21218, USA}





\begin{abstract}

The Prime Extragalactic Areas for Reionization and Lensing Science (PEARLS), a {\jwst} GTO program, obtained a set of unique NIRCam observations that have enabled us to significantly improve the default photometric calibration across both NIRCam modules. The observations consisted of three epochs of 4-band (F150W, F200W, F356W, and F444W) NIRCam imaging in the Spitzer IRAC Dark Field (IDF)\null. The three epochs were six months apart and spanned the full duration of Cycle 1. As the IDF is in the {\jwst} continuous viewing zone, we were able to design the observations such that the two modules of NIRCam, modules A and B, were flipped by 180 degrees and completely overlapped each other's footprints in alternate epochs. We were therefore able to directly compare the photometry of the same objects observed with different modules and detectors, and we found significant photometric residuals up to $\sim$ 0.05 mag in some detectors and filters, for the default version of the calibration files that we used (\texttt{jwst\_1039.pmap}). Moreover, there are multiplicative gradients present in the data obtained in the two long-wavelength bands. The problem is less severe in the data reduced using the latest pmap (\texttt{jwst\_1130.pmap} as of September 2023), but it is still present, and is non-negligible. We provide a recipe to correct for this systematic effect to bring the two modules onto a more consistent calibration, to a photometric precision better than $\sim$ 0.02 mag.

\end{abstract}



\section{Introduction} \label{sec:intro}


For any space-based facilities, pre-determined flux calibrations are essential because the observations of standard sources cannot be integrated into the observations of every science target. For the Near Infrared Camera (NIRCam) \citep{Rieke2005} onboard the James Webb Space Telescope ({\jwst}), the requirement of its absolute flux calibration is to reach an accuracy of 5\% (4\% has been achieved as reported by \citealt{Rigby2023}). The internal uniformity over all its detectors can be much better, and a lot of science applications will benefit from an uniformity of 1--2\%. Over the past year, the NIRCam flux calibration has improved significantly, and the updates have been continuously incorporated in the {\jwst} calibration reference files used by its data reduction pipeline. In the {\jwst} jargon, the reference ``pmap'' files are part of the ``context'' of the evolving pipeline. Due to its importance, the accuracy of the calibrations provided by the pmap has been checked by various independent research groups from time to time \citep[e.g.,][]{Boyer2022,Nardiello2022,Sunnquist2022,Griggio2023,Nardiello2023}, and some of these results have been incorporated in the frequently updated pmap file.


The Prime Extragalactic Areas for Reionization and Lensing Science (PEARLS), a {\jwst} GTO program (PID 1176 \& 2738; \citealt{Windhorst2023}), finished a set of unique NIRCam observations that revealed an unexpected problem in the existing NIRCam flux calibration. We found that the two modules of NIRCam, modules A and B, have non-negligible offsets in their calibrations as provided by the pmap. This paper presents our investigation of this problem and offers a recipe for remedy. The paper is organized as follows: the data and the reduction are described in Section \ref{sec:data}, the analysis method is detailed in Section \ref{sec:phot}, and the results are given in Section \ref{sec:result}.  Throughout the paper, all magnitudes quoted are in the AB system.

\section{Observation, Data and the Revealed Problem} \label{sec:data}

\begin{figure}
    \centering
    \includegraphics[width=0.4\textwidth,trim={3cm 0 0 0}]{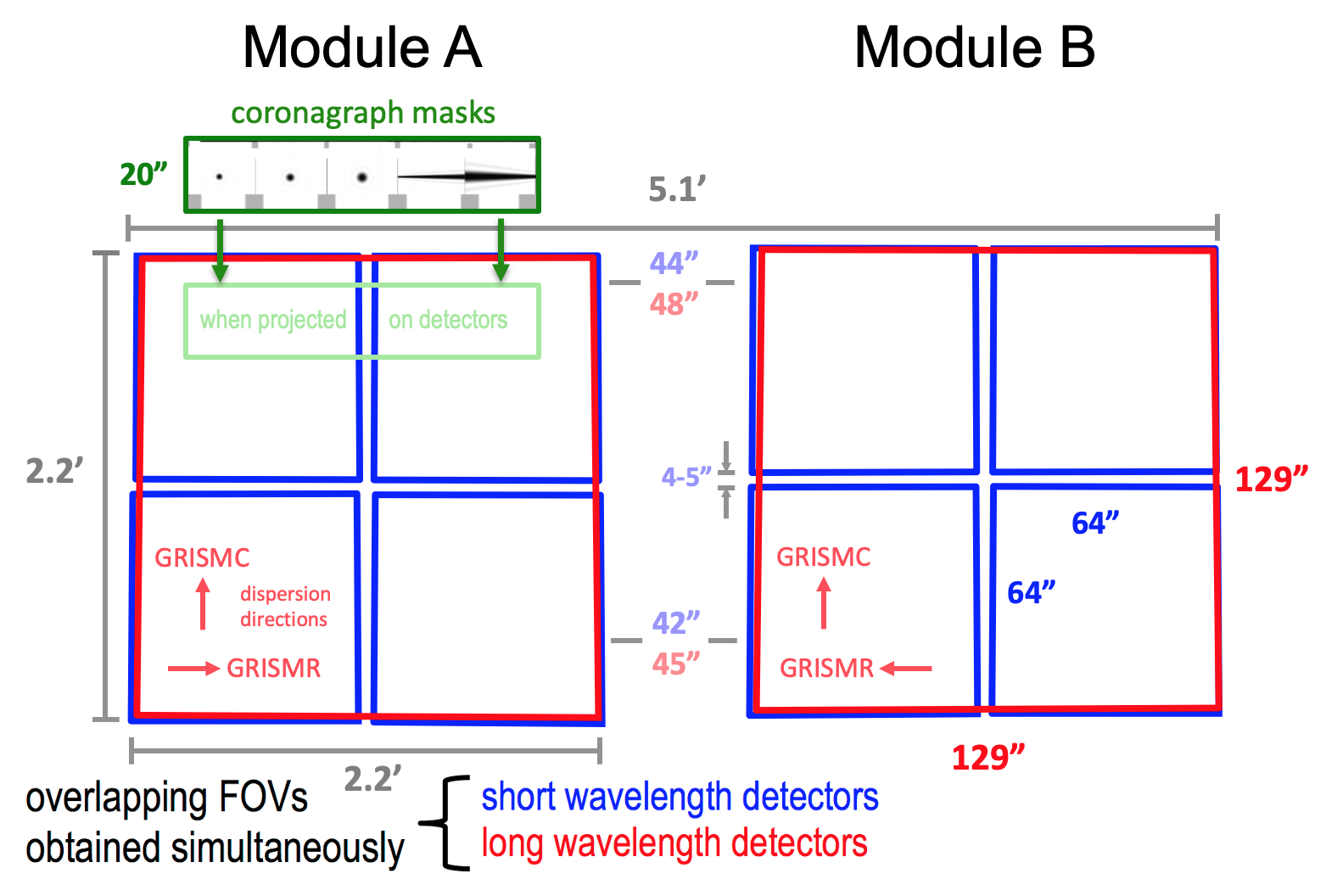}
    \includegraphics[width=0.4\textwidth]{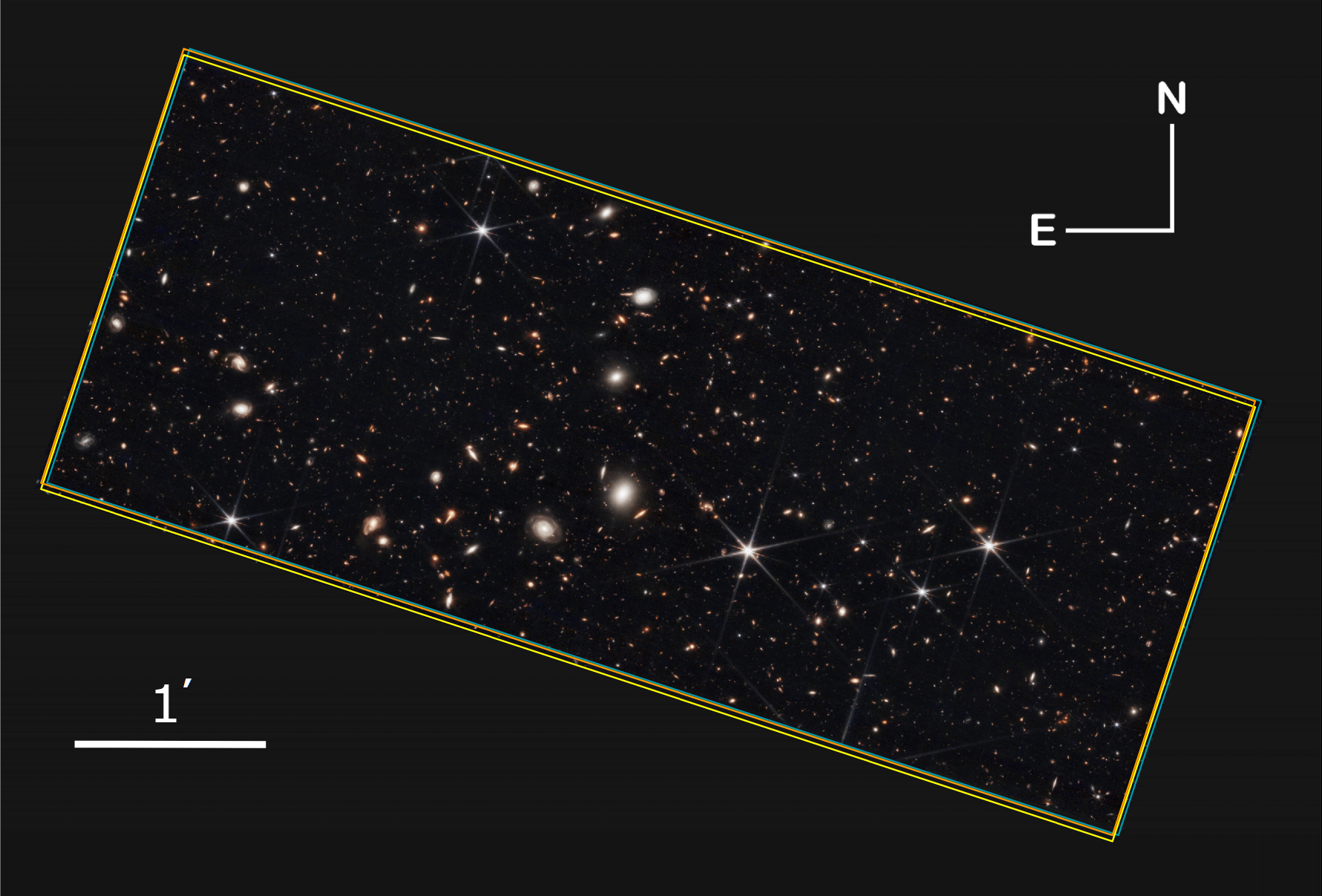}
    \caption{Left: NIRCam focal plane layout (\url{https://jwst-docs.stsci.edu/jwst-near-infrared-camera/nircam-instrumentation/nircam-field-of-view}). Right: Footprints of PEARLS NIRCam observations in the JWIDF. The outlines are for the three epochs, which fully overlapped.}
    \label{fig:idf_footprint}
\end{figure}

The PEARLS NIRCam data on the IRAC Deep Field (dubbed ``JWIDF''; \citealt[]{Yan2023a}) were taken in three epochs six months apart (2022-07-08, 2023-01-06, and 2023-07-06) that spanned the full duration of {\jwst} Cycle 1 for time-domain studies. The first epoch of observations was described by \citet[]{Yan2023a}. In all three epochs, the observations were done in four broad bands, namely, F150W, F200W, F356W, and F444W\null. The first two bands are in the NIRCam ``short wavelength'' (SW) channel, which has four arrays (ID 1--4) in each of the two modules (A and B), and the last two bands are in the ``long wavelength'' (LW) channel, which has one array (ID ``long'') in each module. The detector layout is illustrated in the left panel of Figure~\ref{fig:idf_footprint}.

To cover the gaps between detectors, the observations used FULLBOX dithers with the 6TIGHT pattern, which results in a $\sim5\farcm9\times2\farcm4$ rectangular area covered by six dithered exposures. The footprint is shown in the right panel of Figure~\ref{fig:idf_footprint}. The dithered positions are determined by the STANDARD subpixel dither to optimally sample the point-spread functions (PSFs). For each exposure, the SHALLOW4 readout pattern was adopted with ``up-the-ramp'' fitting to determine the count rate. The observations at each epoch were taken with the same exposure time for all filters,  with comparable exposure times for the different epochs (3157, 2512, 2835)~s for epochs (1, 2, 3).



As the JWIDF is in the {\jwst} northern continuous viewing zone (CVZ), we were able to design the observations such that module A and B were flipped by 180 degrees and completely overlapped each other's footprints in all epochs. Thanks to this unique design, we are able to compare the fluxes of the same objects measured in different arrays of different modules.

We create the mosaics (one for each band) following the procedures outlined by \citet[][]{Yan2023b}. In brief, the \texttt{uncal} data are downloaded from the MAST archive, and are processed by the {\jwst} pipeline\footnote{\url{https://github.com/spacetelescope/jwst}} \citep[version 1.11.4;][]{Bushouse2022} stages \texttt{calwebb\_detector1} and \texttt{calwebb\_image2} to produce the \texttt{cal} data.
This involves a few customized steps. Before producing the \texttt{cal} data, for each image in the SW channel, we estimate and subtract the median background count rate to level off the different baseline bias levels among SW detectors. To do this, each source-masked image is segmented into blocks of 128$\times$128 pixels, and then we apply a 3$\times$3 median filter to obtain the median background value. For the produced \texttt{cal} images, we correct for the 1/f noise in each image in the SW channel by running the external tool \texttt{image1overf}\footnote{\url{https://github.com/chriswillott/jwst}} on a per-amplifier basis along both rows and columns. Then, we apply another round of background subtraction on each \texttt{cal} image, which aims to remove the non-uniformity of the background in the final mosaic and is also done by segmenting each source-masked image into sections of 128$\times$128 pixels and applying a 3$\times$3 median filter. The processed \texttt{cal} data are then fed to the {\jwst} pipeline stage \texttt{calwebb\_image3} to create the mosaics. All mosaics are reduced to the same World Coordinate System (WCS) grid using the epoch 1 data as the reference to allow for matched aperture photometry (see section \ref{sec:phot}). The final images have a pixel scale of 0\farcs06, and the absolute astrometry is tied to the GAIA third data release.

\begin{figure}
    \centering
    \includegraphics[width=0.9\textwidth]{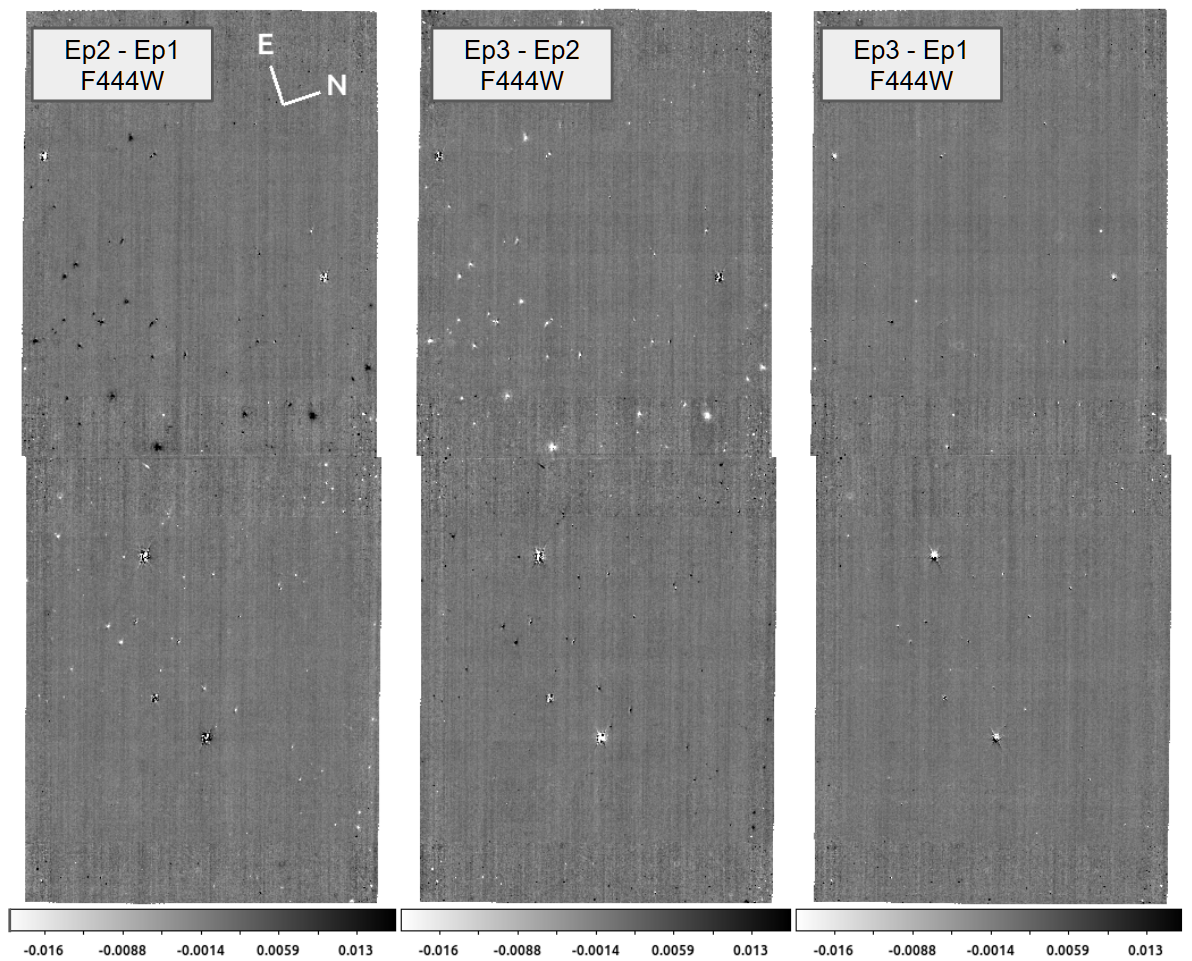}
    \caption{Demonstration of the over/under-subtraction problem in the difference images between the three JWIDF epochs using the default calibration. The F444W band is used here as the effect is the most obvious in this band. The parent images are produced using the latest \texttt{jwst\_1130.pmap}. From left to right, the difference images are between epochs 2 and 1, 3 and 2, and 3 and 1, respectively. These images are smoothed using a circular aperture of 6 pixels in radius to make the effect more visible in this display. The problem only appears when the epoch 2 data are involved. As the NIRCam detector orientation was flipped by 180\arcdeg~in epoch 2 with respect to that in epoch 1 or 3, the only plausible explanation is that the flux calibrations in module A and B have non-negligible offset.
}
    \label{fig:sub_before}
\end{figure}

When constructing difference images in between epochs for transient search, however, an unexpected problem was revealed. This is demonstrated in Figure \ref{fig:sub_before}. The difference images between epochs 2 and 1 show under-subtraction features in half of the field and over-subtraction features in the other half. This is also seen in the difference images between epochs 3 and 2. However, the difference images between epochs 3 and 1 do not have this problem.

\begin{figure}
    \centering
    \includegraphics[width=0.5\textwidth]{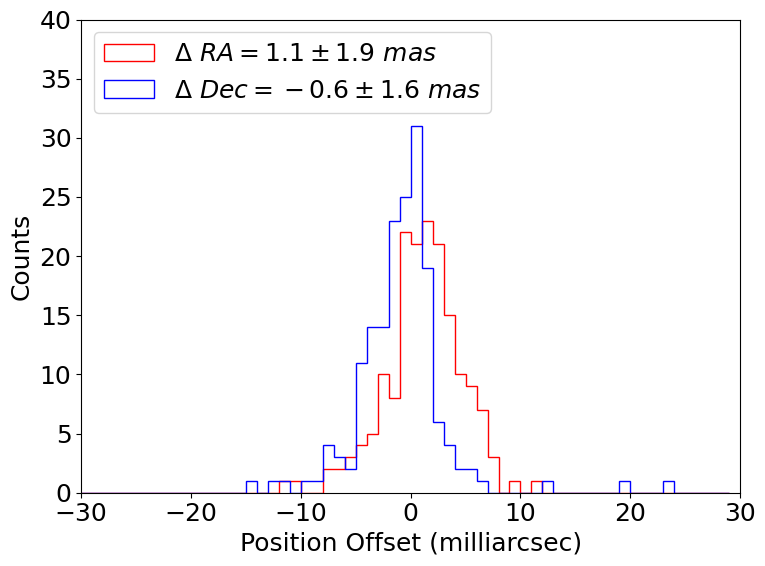}
    \caption{Histograms showing position offsets of images between epochs 1 and 2, measured using bright ($<22$~mag) sources of the JWIDF field in the F356W band. The red histogram is for the RA direction, while the blue one is for the Dec direction. The values and uncertainties quoted in the figure legend are the median and median absolute deviation (MAD) values of the measured offsets.}
    \label{fig:astrometry}
\end{figure}

The usual cause of under/over-subtraction feature is imperfect image alignment. However, this is not the case here because the internal alignment accuracy of all our images has reached $\lesssim 2$ milli-arcsecond (mas). This is demonstrated in Figure~\ref{fig:astrometry}, which shows the histogram of the measured position offsets of bright  ($< 22$~mag), unsaturated, compact sources in F356W (as an example) between epoch 1 and 2. The median and median absolute deviation (MAD) values along the RA and Dec directions are $1.1\pm 1.9$ and $0.7\pm2.2$~mas, respectively, which are much smaller than the pixel size of 60~mas. In Table~\ref{tab:astrometry}, we list the measured median and MAD position offsets for all the 4 bands. The result suggests that the problem cannot be attributed to image misalignment. Noticing that the problem only occurs when the epoch 2 data are involved, and considering that the orientation of module A and B was the same in epoch 1 and 3 but was flipped by 180{\arcdeg}~in epoch 2, we are forced to consider the possibility that the flux calibrations of the two modules have offsets.

\begin{table}[ht]
    \centering
    \begin{tabular}{ccccc}
    \hline\hline
    Band & $N_{obj}$ & $\Delta$ RA (mas) & $\Delta$ Dec (mas) \\
    \hline
    F150W & 171 & -0.6 $\pm$ 2.8 & -0.9 $\pm$ 2.8 \\
    F200W & 213 & 1.0 $\pm$ 2.3 & -1.5 $\pm$ 3.1 \\
    F356W & 171 & 1.1 $\pm$ 1.9 & -0.6 $\pm$ 1.6  \\ 
    F444W & 158 & 0.7 $\pm$ 2.2 & 0.4 $\pm$ 2.5 \\
    \hline
    \end{tabular}
    \caption{Median and MAD of the position offsets between sources detected in epochs 1 and 2.}
    \label{tab:astrometry}
\end{table}

\section{Analysis} \label{sec:phot}

\subsection{Detector grouping}

For ease of the analysis, we separate the raw data into groups by epoch, band, and detector array. Table~\ref{tab:data_groups} summarises the number of raw images in each band in a given epoch. (All three epochs have the same scheme.)

\begin{table}[ht]
    \centering
    \begin{tabular}{c|cccc|cccc}
        \hline\hline
        Array Name & \multicolumn{4}{c|}{Number of Raw Images} & \multicolumn{4}{c}{Number of Sources Used}  \\
                    &   F150W & F200W & F356W & F444W   &   F150W & F200W & F356W & F444W       \\
        \hline
        nrca1 & 6 & 6 & \nodata & \nodata & 19 & 23 & \nodata & \nodata  \\
        nrca2 & 6 & 6 & \nodata  & \nodata & 29 & 34 & \nodata & \nodata  \\
        nrca3 & 6 & 6 & \nodata  & \nodata & 28 & 31 & \nodata & \nodata  \\
        nrca4 & 6 & 6 & \nodata  & \nodata  & 32 & 34 & \nodata & \nodata \\
        nrcalong & \nodata  & \nodata  & 6 & 6 & \nodata & \nodata & 97 & 84 \\
        nrcb1 & 6 & 6 & \nodata  & \nodata & 19 & 21 & \nodata & \nodata  \\
        nrcb2 & 6 & 6 & \nodata  & \nodata & 19 & 23 & \nodata & \nodata  \\
        nrcb3 & 6 & 6 & \nodata  & \nodata  & 26 & 30 & \nodata & \nodata \\
        nrcb4 & 6 & 6 & \nodata  & \nodata  & 25 & 31 & \nodata & \nodata  \\
        nrcblong & \nodata  &  \nodata & 6 & 6 & \nodata & \nodata & 83 & 82  \\       
        \hline
    \end{tabular}
    \caption{Summary of the image groups for making mosaics per detector array per epoch, and the number of sources used for the analysis in each mosaic.}
    \label{tab:data_groups}
\end{table}

For each group, the mosaics are created in a similar way as described above.
As the {\jwst} calibration context file has been continuously updated in Cycle 1, the adoption of different context files will result in different flux calibrations. For this work, we created two sets of mosaics with two contexts, namely, \texttt{jwst\_1039.pmap} (published on 2023-01-12) and \texttt{jwst\_1130.pmap} (published on 2023-09-15). The latter is the latest context as of this writing.

\begin{figure}[ht]
    \centering
    \includegraphics[width=0.5\textwidth]{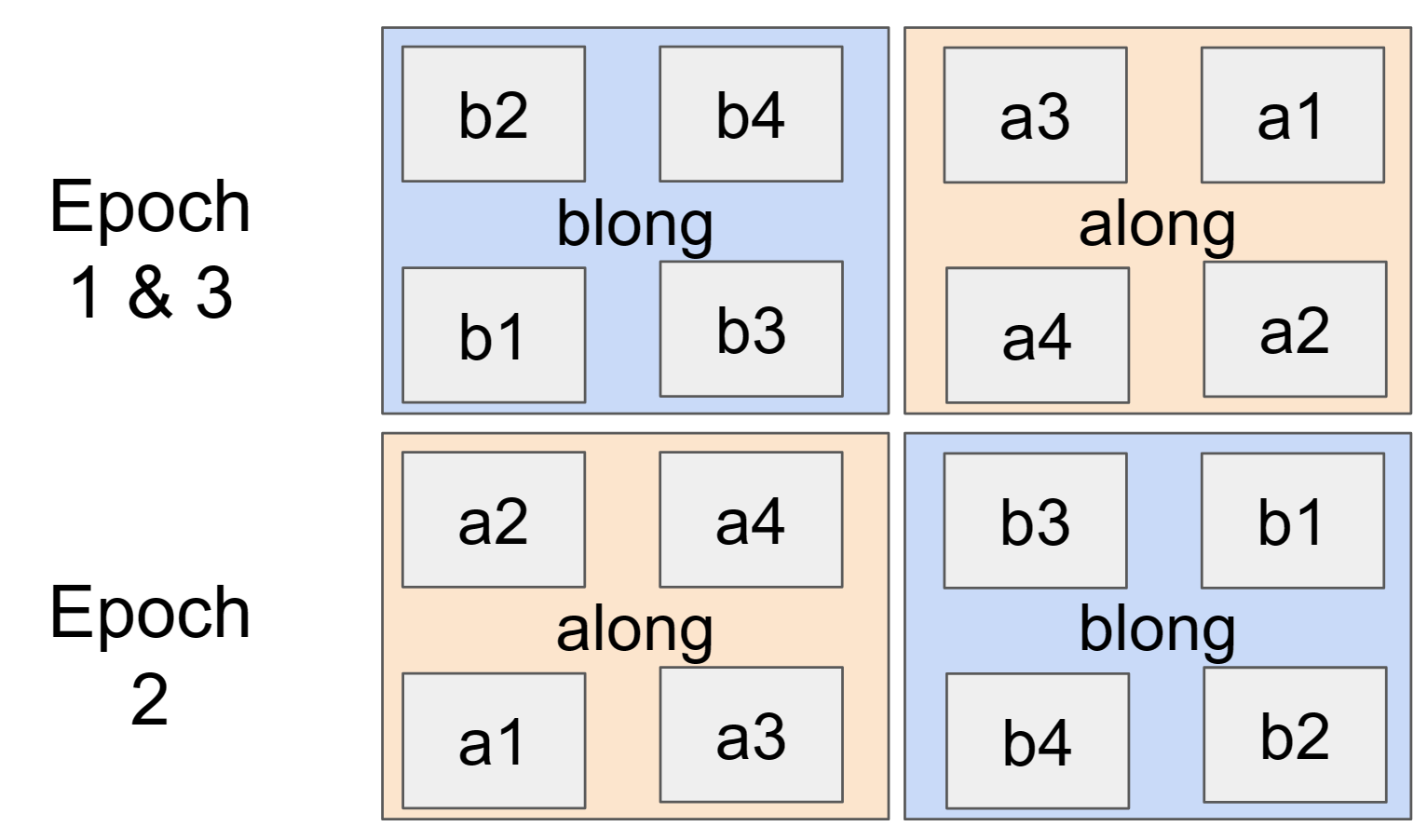}
    \caption{PEARLS JWIDF observation scheme. By design, the position angles of the three epochs differ by 180\arcdeg, which means that the same patch of sky (hence the same set of objects) are observed independently by different arrays of the two different modules.}
    \label{fig:obs_scheme}
\end{figure}

By design, the position angles of two consecutive epoch JWIDF observations differ by 180\arcdeg, which means that the same patch of sky (hence the same set of objects) are observed independently by two different arrays from the two modules. Moreover, because of the way the arrays are labeled (center-symmetric), the matching pair of arrays have the same ID (1, 2, 3, 4, or ``long''). As illustrated in Figure~\ref{fig:obs_scheme}, epochs 1 and 3 have the same projected array layout on the sky, whereas between epochs 1 and 2 (2 and 3), the same patch on the north-east half of the field is observed by modules B and A (A and B), respectively, and it is the other way around for the southwest half. 

For each patch of sky in each band, we perform matched aperture photometry using the \texttt{SExtractor} \citep{Bertin1996} dual-image mode on the three mosaics corresponding to the three epochs, with the first epoch mosaic used as the detection image. We adopt \texttt{MAG\_ISO} as the magnitude for all our analyses. Using other flux metrics such as \texttt{MAG\_AUTO} does not change the result. To ensure the robustness of the statistics, we only include the objects that are brighter than 22~mag, which corresponds to signal-to-noise ratio (SNR) $\gtrsim$~100. The number of sources used for the analysis are listed in Table~\ref{tab:data_groups}.

\subsection{Modeling}\label{sec:model}

The measured magnitudes of a source can be written as
\begin{align}
    m = m_\mathrm{source} + m_\mathrm{sys} + \sigma \,,
\end{align}
where $m_{source}$ is the true source magnitude, $m_\mathrm{sys}$ is the systematic error, and $\sigma$ is the random error. For a given pmap, both $m_\mathrm{sys}$ and $\sigma$ can depend on epoch and/or the detector array that the source falls in. Using $m_\mathrm{array}$ and $m_\mathrm{time}$ to denote the systematic error terms associated with the detector array and the time of observation, respectively, we have

\begin{align}
    m = m_\mathrm{source} + m_\mathrm{array} + m_\mathrm{time} + \sigma \,.
\end{align}

The magnitude offsets measured between any two epochs can then be expressed as:
\begin{align}
    \Delta & = \Delta_\mathrm{source} + \Delta_\mathrm{array} + \Delta_\mathrm{time} + \sigma \, ,
\end{align}
where $\Delta_\mathrm{source}$ is the intrinsic change of the source flux, $\Delta_\mathrm{array}$ is the inter-module calibration systematic error, and $\Delta_\mathrm{time}$ is the systematic error related to the time of observation.  When combining (e.g., using mean or median) all the sources in a patch of sky, the random error term $\sigma$ should disappear. Furthermore, for the JWIDF data, we expect a negligible amount of intrinsically variable sources in each sky patch, i.e., $\Delta_\mathrm{source} \simeq 0$. With all this, we can reduce the above equation to 
\begin{align}
    \Delta & \simeq \Delta_\mathrm{array} + \Delta_\mathrm{time} \,.
\end{align}

Due to the 180\arcdeg~rotation of the field of view, between two successive epochs $T_i$ and $T_j$, we have the pair of arrays with the same ID $x$, $x \in \{1, 2, 3, 4, long\}$, from modules A and B measured in two different sky patches $p$ and $q$ but in flipped time order. This is expressed as (random error term omitted):

\begin{align}
    \Delta_p & = \Delta_{A_x, B_x} + \Delta_{T_i, T_j} \, , \label{eq:sysp} \\
    \Delta_q & = \Delta_{B_x, A_x} + \Delta_{T_i, T_j} \, , \label{eq:sysq}
\end{align}
where, by definition, $\Delta_{A_x, B_x} = - \Delta_{B_x, A_x}$. Therefore, we can separate the two systematic error terms: 
\begin{align}
    \Delta_{A_x,B_x} & = (\Delta_p - \Delta_q) /2 \, , \\
    \Delta_{T_i,T_j} & = (\Delta_p + \Delta_q) /2 \, .\label{eq:syst}
\end{align}

\section{Result} \label{sec:result}

\begin{figure}
    \centering
    \includegraphics[width=0.9\textwidth]{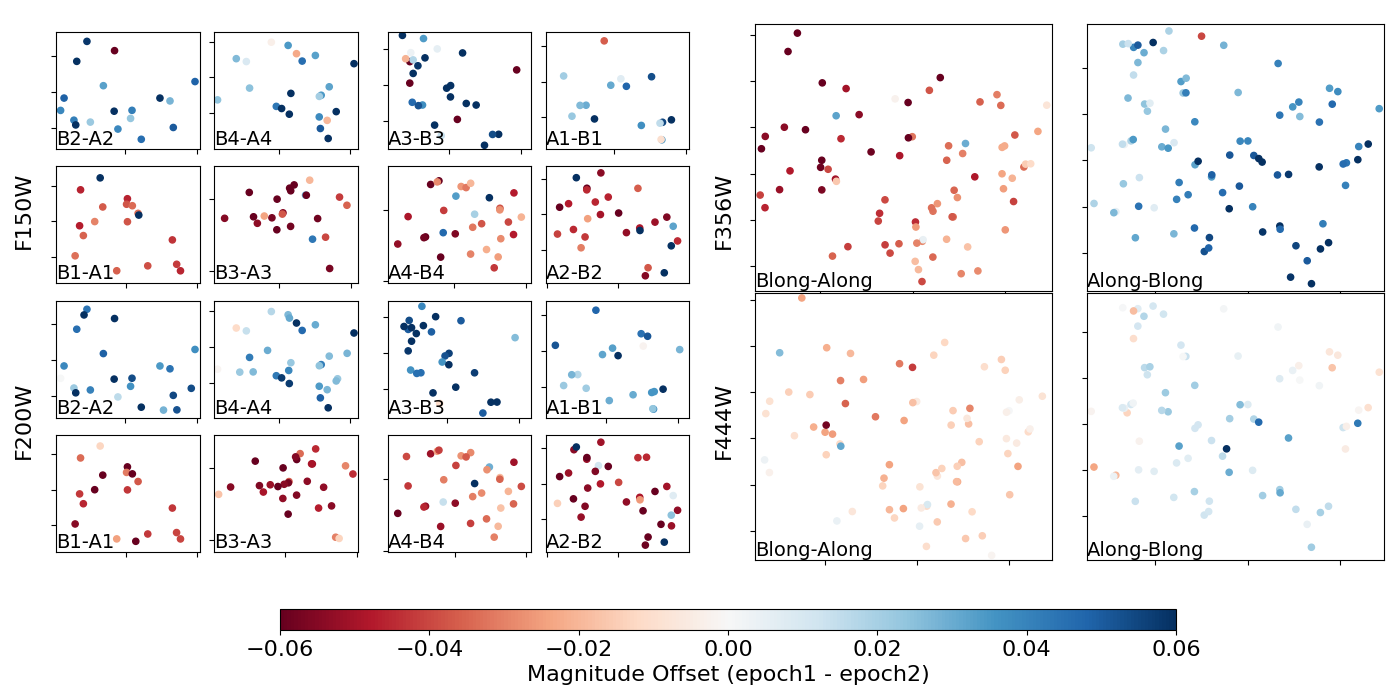}
    \includegraphics[width=0.9\textwidth]{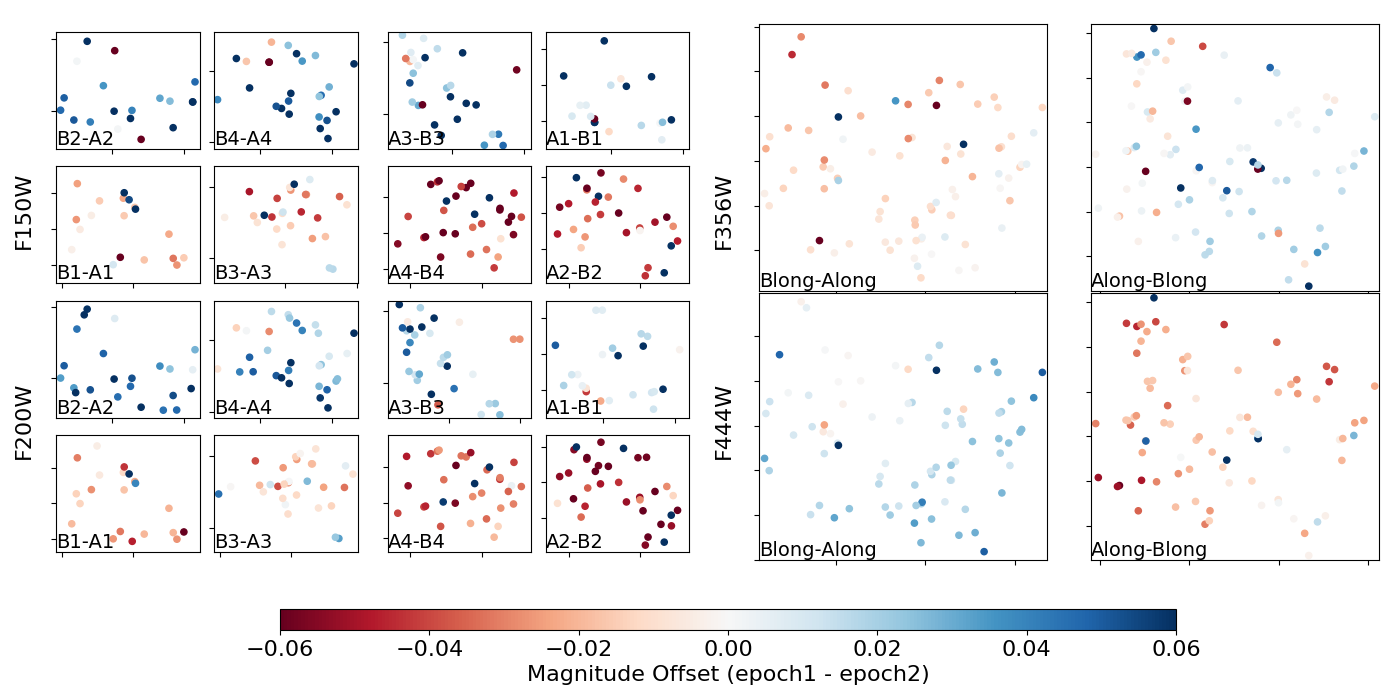}
    \caption{Sources detected in each detector array and each band. The color coding indicates the measured magnitude offsets between epochs 1 and 2. Top: results derived from \texttt{jwst\_1039.pmap}; Bottom: results derived from \texttt{jwst\_1130.pmap}}.
    \label{fig:xy_dmag}
\end{figure}

\begin{figure}
    \centering
    \includegraphics[width=0.45\linewidth]{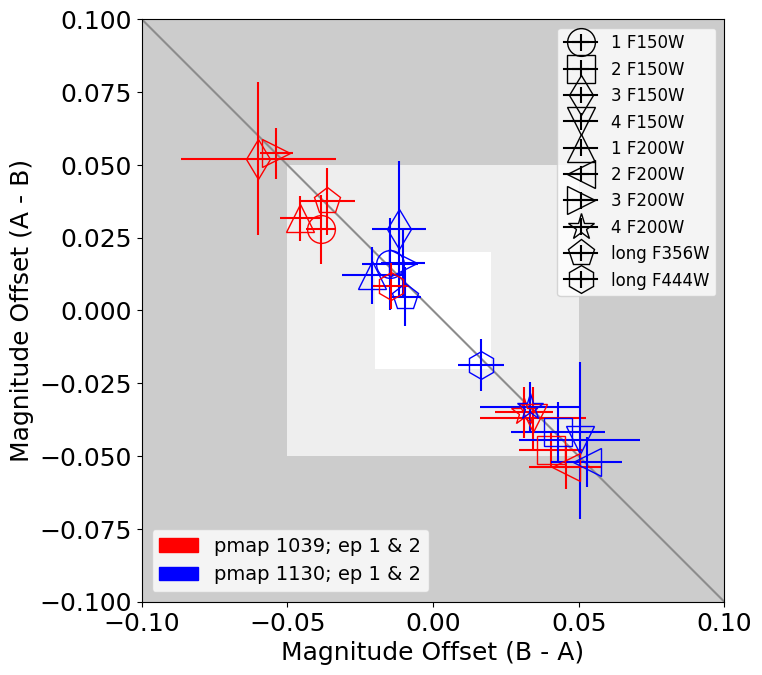}
    \includegraphics[width=0.4\linewidth]{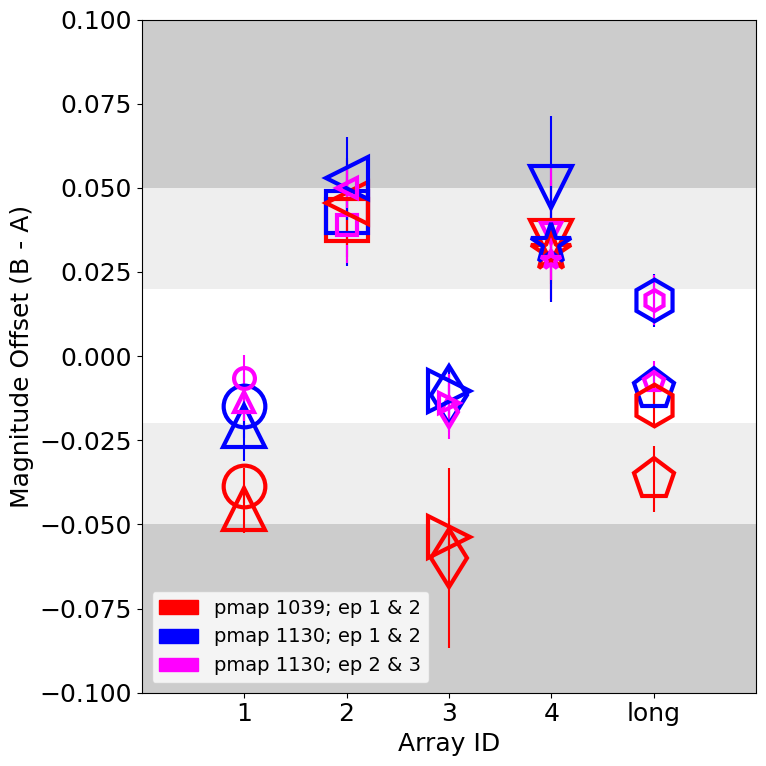}
    \caption{Left: Median magnitude offsets in between epochs 1 and 2 for all detectors. The error bars are the MAD values. Different symbols are for different bands as indicated in the legends. The ones in red are from the data processed by \texttt{jwst\_1039.pmap}, while the blue ones are from those processed by \texttt{jwst\_1130.pmap}. Each data point represents the combined measurements from the pair of arrays with the same ID; each such pair is observed twice in two patches of sky. The x-axis is from the sources observed by module B in epoch 1 and by module A in epoch 2, and y-axis is from the sources observed by module A in epoch 1 and by module B in the epoch 2. This resembles the $\Delta_p$ and $\Delta_q$ terms in Equation~\ref{eq:sysp} and \ref{eq:sysq}. The black line indicates the location where $\Delta_p + \Delta_q = 0$. The two shaded regions in different grey scales indicate the offsets of 0.05 and 0.02~mag (corresponding to $\sim5\%$ and $\sim2\%$ relative error, respectively), used as a visual guide. If the flux calibration has no detector-dependent offset, all the points should be at the center (0, 0). However, this is not the case, even for those that are based on the latest pmap (the blue points).
    Right: Similar to the left panel, but in an alternative presentation. In addition to the data points showing the epoch 1 and 2 differences based on two different pmaps files (red and blue symbols), a set of magenta data points (rendered smaller in size for better clarity) are added to show the epoch 2 and 3 differences based on \texttt{jwst\_1130.pmap}. The x-axis is the array ID, and y-axis is the measurements from the sources observed by module B in epoch 1 (and epoch 3 as well), which were observed by module A in the second epoch. The magenta points and the blue points, both based on the data processed by \texttt{jwst\_1130.pmap}, are very close to each other because the sources were observed by the same detectors.
}
    \label{fig:summary_dmag}
\end{figure}

The measured magnitude differences are shown in Figure~\ref{fig:xy_dmag}, which shows that the magnitude offsets between epochs 1 and 2, measured from the same sources, depends on which detector array observed the source. 

Taking the \texttt{jwst\_1039.pmap} F356W image (the top-right two panels) as an example, the left side (north-east half of the JWIDF) is first observed with module B in epoch 1, and then module A in epoch2, and it is the reverse for the right side (south-west half of the JWIDF). The red colors shown in the left panel indicate that the sources measured in module B (epoch 1) are $\sim$0.04~mag  brighter than module A (epoch 2), while the blue colors shown in the right panel indicate that the sources measured in module A (epoch 1) are around the same amount (i.e, $\sim$0.04~mag) dimmer than module B (epoch 2). The same is true for the other bands; in particular, for the SW modules, array ID 1 and 3 have the module B sources brighter, but in array ID 2 and 4, the sources are brighter in module A.

Comparing the results based on \texttt{jwst\_1039.pmap} and \texttt{jwst\_1130.pmap}, we see that in general, the magnitude offsets are smaller when using the latter pmap, and more so in some arrays than others. Another notable feature is that, in both cases, there are obvious gradients in the LW bands, which are indicated by the fading colors of points along the long edge of the JWIDF field in the panels to the right in Figure~\ref{fig:xy_dmag}.

To best quantify the measured magnitude offsets as the systematic error associated with the module/array/epoch, we calculate the median and median absolute deviation (MAD) of the offsets in each sky patch, and organize the data points based on the notion outlined in Section~\ref{sec:model}. The result is shown in Figure~\ref{fig:summary_dmag}.

In the left panel, each data point represents the combined measurements from the pair of arrays with the same ID (1, 2, 3, 4, or ``long''), as each such pair is observed twice in two patches of sky in one band. The x-axis is from the sources observed by module B in the first epoch, and y-axis is from the sources observed by module A in the first epoch. This resembles the $\Delta_p$ and $\Delta_q$ terms in Equation~\ref{eq:sysp} and \ref{eq:sysq}. The black line indicates the location where $\Delta_p + \Delta_q = 0$. In the right panel, we plot the median magnitude offsets as the y-axis, but in the x-axis, it is the detector array ID, for better comparing the results on a per-array basis.

The results show that:
\begin{itemize}
    \item There are large offsets in the inter-module calibration from the data reduced in both pmaps. The offsets can be $\gtrsim$ 0.05 mag in some arrays and bands. From the right panel, we can see that the newer pmap (blue and magenta symbols) have improved calibration ($\lesssim$ 0.02 mag) for the detector array ID 1 and 3, and maybe also for array ID ``long''; however, for array ID 2 and 4, the offsets remain the same as the older pmap.
    \item The data show little or no time-dependent systematic errors in the calibration. This is suggested in the left panel because all data points fall close to the black line where $\Delta_p + \Delta_q = 0$. According to Equation~\ref{eq:syst}, this is consistent with the scenario where no systematic error is associated with the time of observation. The same conclusion can also be derived from comparing the magenta data points to the blue ones on the right panel. The magenta data points are the measurements from using epoch 3 data in place of the epoch 1 data as used in deriving the blue data points. The fact that they all mostly overlap with each other means that the calibration is consistent between epochs 1 and 3, where the field of view has rotated by 360 degrees thus the same patch of sky is observed by the same array and therefore the array-dependent systematic error term does not play a role.
\end{itemize}

\section{Discussion}


\subsection{Linear model fitting to the measured source magnitude offsets}\label{sec:linear_fit}

In the previous section we derive the median magnitude offsets between modules A and B, which would only be appropriate to represent the calibration systematic error related to the detector arrays if they are not a function of the location of the sources on the sky. However, this is not true for the LW bands (F356W and F444W), where we can see gradients across the long-side direction of JWIDF. Furthermore, the result suggests that there is little to no time-dependent calibration error, which imposes a strong constraint that ties the pair of sky patches that are observed by the same array ID together. This can be exploited to derive more robust estimates of the inter-module systematic errors.

To this end, we construct the following linear model to fit the magnitude offsets measured from both patches of sky (i.e., $\Delta_p$ and $\Delta_q$) that are observed by two arrays of the same array ID but in reversed time order:

\begin{align}
    \Delta_p &= kd + \Delta_{A,B} \\
    \Delta_q &= kd + \Delta_{B,A} = kd - \Delta_{A,B} \, , \label{eq:linear_model}
\end{align}
where $d$ is the 1-d coordinate of a source along the direction of the long-side of JWIDF ($\sim$ 72{\arcdeg} from North C.C.W.),  $k$ and $\Delta_{A,B}$ are the linear model parameters, being the slope of the gradient and the inter-module calibration magnitude offset.

The model fitting is done for all pairs of sky patches. For the SW bands (F150W and F200W), we force the slope to be 0. The best-fit model parameters and the uncertainties derived from the \texttt{jwst\_1130.pmap} mosaics are listed in Table~\ref{tab:mag_off_correction}. The fitted magnitude offsets between module A and module B are consistent with the median values reported in \S~\ref{sec:result} for the SW bands, but the values are much smaller for the LW bands. This is reconciled by the fact that the fitted slope of the gradients presented in F356W and F444W bands are $\sim0.03$~mag~arcmin$^{-1}$, which could lead to the seemingly larger values derived from taking the median of the samples.

\begin{table}[ht]
    \centering
    \begin{tabular}{cccc}

\hline\hline
Array ID & Filter & $\Delta_{A,B}$ & Slope (mag~arcmin$^{-1}$) \\
\hline
1 & F150W & 0.01296 $\pm$ 0.00155 & \nodata \\
1 & F200W & 0.01534 $\pm$ 0.00111 & \nodata \\
2 & F150W & -0.04342 $\pm$ 0.00148  & \nodata \\
2 & F200W & -0.05267 $\pm$ 0.00117  & \nodata \\
3 & F150W & 0.02193 $\pm$ 0.00228 & \nodata \\
3 & F200W & 0.01397 $\pm$ 0.00113  & \nodata \\
4 & F150W & -0.05013 $\pm$ 0.00231  & \nodata \\
4 & F200W & -0.03045 $\pm$ 0.00158  & \nodata \\
long & F356W & 0.00874 $\pm$ 0.00110 & 0.00300 $\pm$ 0.00075 \\
long & F444W & -0.01654 $\pm$ 0.00128  & 0.00433 $\pm$ 0.00090 \\
\hline
    \end{tabular}
    \caption{Best fit linear model parameters derived from the \texttt{jwst\_1130.pmap} mosaics.}
    \label{tab:mag_off_correction}
\end{table}

\subsection{Recipe for correction}

The best-fit models derived from \S~\ref{sec:linear_fit} can be used to correct for this intra-module calibration systematic error. To do this, we need to decide a ``baseline'' or origin for which we will attribute the measured $\Delta_{A,B}$ value across the two modules A and B. Comparing the photometry from the images reduced using the two pmaps \texttt{jwst\_1039.pmap} and \texttt{jwst\_1130.pmap}, we notice that for the sources observed by array ID 1 and 3, where we see improved calibration consistency between the two modules, the module B photometry remains the same while the module A photometry changes from the older pmap to the newer one.
Therefore, we adopt the photometry calibration of module B as the baseline and apply the magnitude offset to module A to correct for the effect.

The magnitude offset to be applied to the raw ``cal.fits'' files can be expressed as:

\begin{align}
    \Delta_A &= \Delta_{A,B} \\
    \Delta_B &= 0 
\end{align}

For the LW bands where gradients are present, one needs to render the slope image with the best-fit slope values (also taking into account for necessary coordinates transformation to match to the definition of $d$ variable used in Equation~\ref{eq:linear_model}), and then divide this slope image from the ``cal.fits'' files.

\begin{figure}[t]
    \centering
    \includegraphics[width=0.9\textwidth]{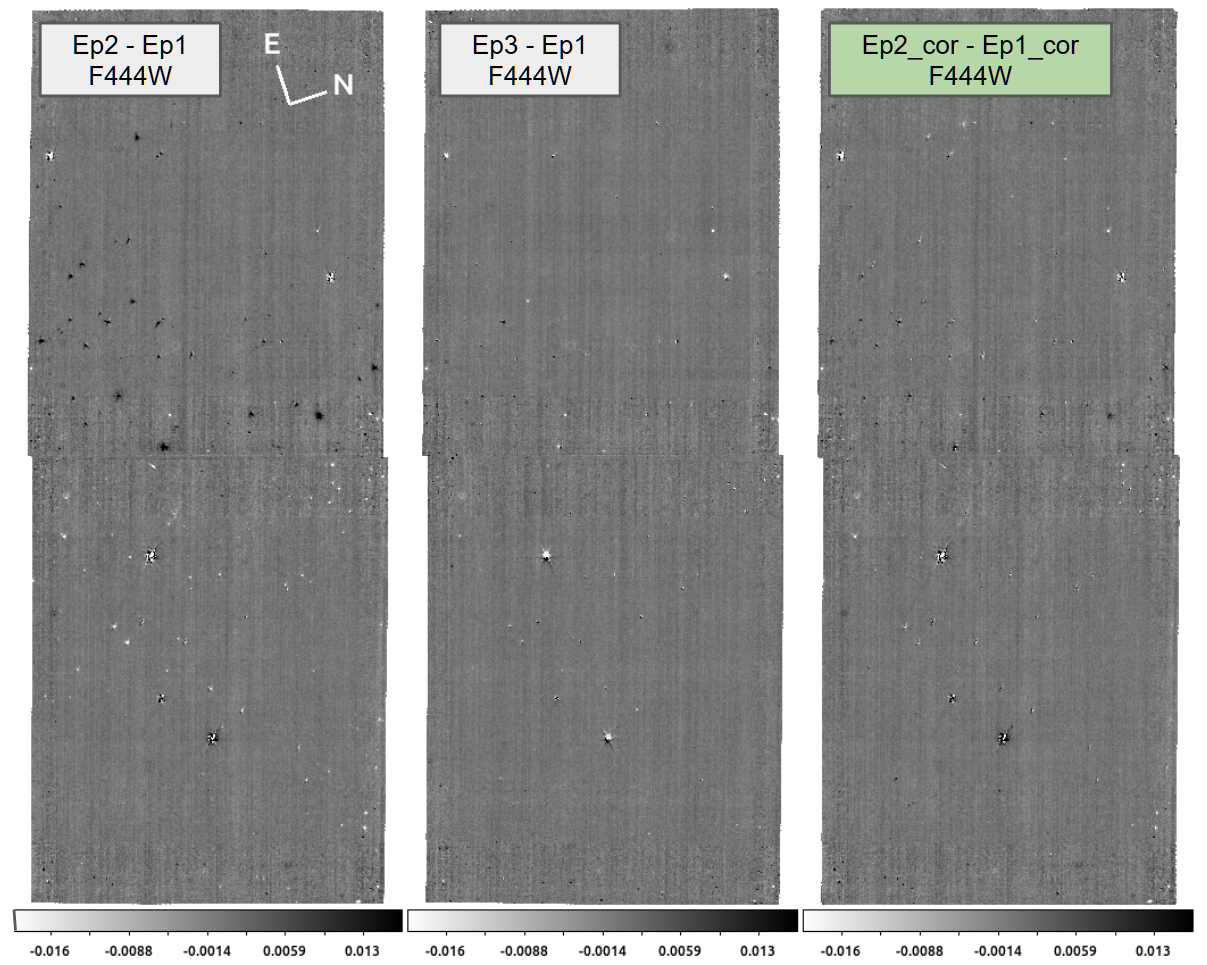}
    \caption{Demonstration of the effectiveness of our correction of the detector-dependent flux calibration non-uniformity. As in Figure \ref{fig:sub_before}, F444W band is used and the images are smoothed for this illustration purpose. The left panel shows the difference image between epochs 1 and 2 where the over/under-subtraction features are prominent (due to the flipped orientation of modules A and B), while the middle panel shows the one between epochs 1 and 3 where few features are seen (due to the same orientation of modules A and B). The right panel again shows the difference image between epoch 1 and 2, however the parent images are created after applying out recipe to correct for the calibration non-uniformity. This difference image is as clean as the one in the middle panel, which shows the improvement brought by our correction procedure.
    }
    \label{fig:sub_after}
\end{figure}

\begin{figure}[t]
    \centering
    \includegraphics[width=0.45\linewidth]{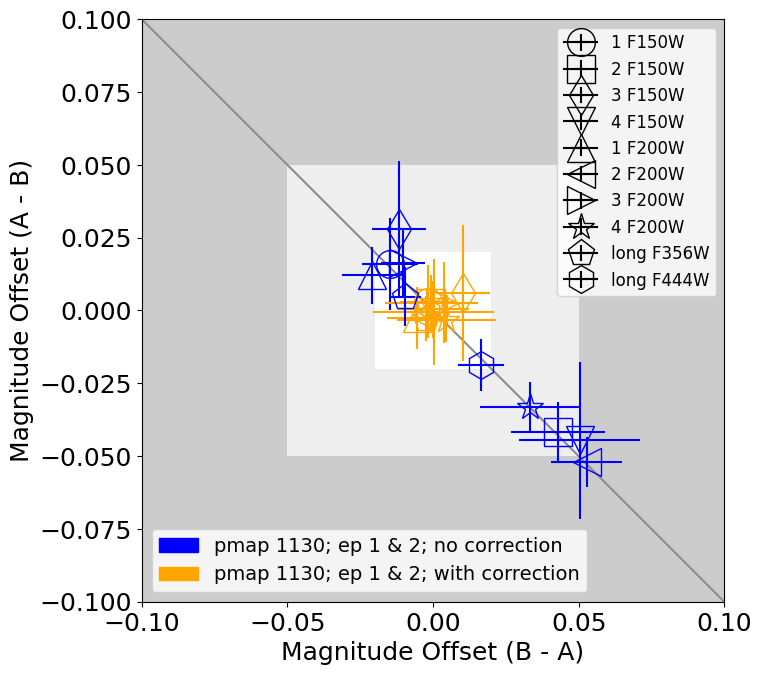}
    \includegraphics[width=0.4\linewidth]{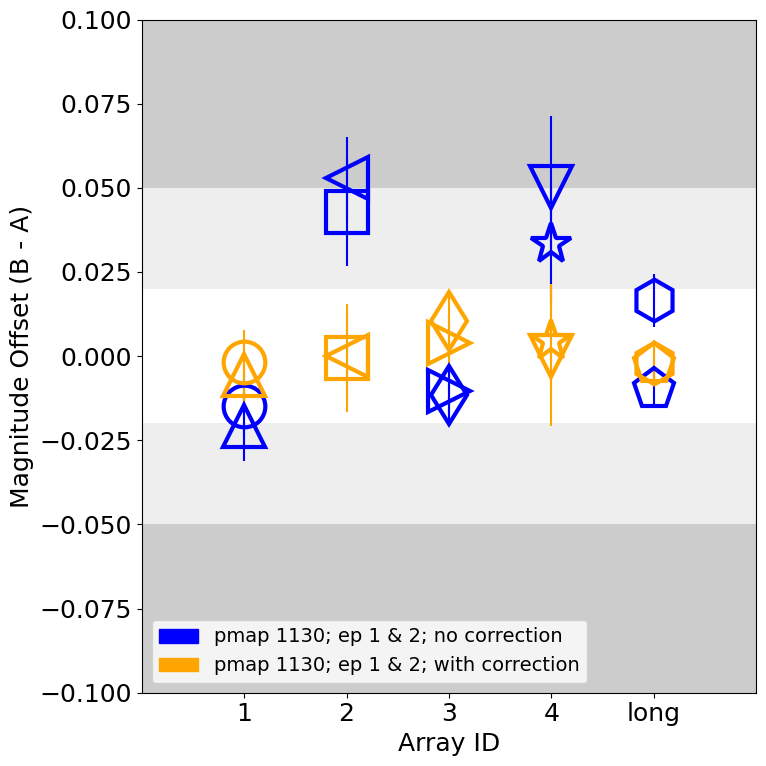}
    \caption{Similar to Figure~\ref{fig:summary_dmag}, but showing comparisons of the median magnitude offsets before and after our detector-dependent corrections to the flux calibrations. In both panels, the blue symbols are the measurements from epochs 1 and 2 based on the data processed using \texttt{jwst\_1130.pmap}, and the orange ones are those made from the same data but with our corrections applied to the \texttt{cal.fits} files. The two shaded regions indicate the offsets of 0.05 and 0.02~mag as in Figure~\ref{fig:summary_dmag}. The orange symbols are all within 0.02~mag to the center (left panel) or the central line (right panel), which shows the improvements made by our corrections.
    }
    \label{fig:summary_dmag_after}
\end{figure}

We tested the recipe on the JWIDF data and created mosaics that are free of this calibration systematic error. The difference images created after the correction do not have the over/under-subtraction features across the two halves of the JWIDF field, as illustrated in Figure~\ref{fig:sub_after}. We run the same analysis as outlined in \S~\ref{sec:phot} with the corrected mosaics and the result is presented in Figure~\ref{fig:summary_dmag_after}. It shows that after the correction, the flux calibrations of modules A and B are consistent, and the offsets are reduced to $<0.02$~mag for all arrays.


\section{Conclusion}

This work exploited the unique design of the three-epoch PEARLS JWIDF observations to characterize the array-dependent calibration systematic errors in the \textit{{\jwst}} NIRCam instrument.
%
While there are little to no systematics associated with the time of observation in the measured magnitudes, the systematics associated with the module/array can be as large as $\sim$0.05~mag. The exact size of the systematic offset depends on the detector array and band. The most recent pmap (\texttt{jwst\_1130.pmap} as of this writing) has improved inter-module calibration for array ID 1 and 3 and maybe also for ``long'', but little improvement is seen in array ID 2 and 4. The offsets in the LW bands show spatial gradients, which are not seen in the SW bands.

We constructed a linear model to describe the behavior of this flux calibration systematic error and fit this model to our measurements. The best-fit parameters are used as the input to our recipe to correct for the effect to bring the two modules onto the same calibration. With our correction, the offset is reduced to $<0.02$~mag. A Python implementation of the recipe is available at the GitHub repository \url{https://github.com/Jerry-Ma/idf_nircam_cal_recipe}. This recipe is only for the four NIRCam bands through which the JWIDF data were taken. It is likely that this problem also presents in other NIRCam bands and can be corrected in the same way, but the exact solutions will have to wait until similar observations in these other bands are obtained, e.g., through a long-term calibration program that uses different detectors observing the same area with the full suite of filters. 



\section*{Acknowledgments}

We dedicate this paper to the memory of our dear PEARLS colleague Mario Nonino, who was a gifted and dedicated scientist, and a generous person. This work is based on observations made with the NASA/ESA/CSA James Webb Space
Telescope. The data were obtained from the Mikulski Archive for Space Telescopes at the Space Telescope Science Institute, which is operated by the Association of Universities for Research in Astronomy, Inc., under NASA contract NAS 5-03127 for {\jwst}. These observations are associated with {\jwst} programs 1176 and 2738.
ZM is supported by the National Science Foundation grants No. 1636621, 2034318, and 2307448.
H.Y. and B.S. acknowledge the partial support from the University of Missouri Research Council Grant URC-23-029 and HST GO-17154.
RAW, SHC, and RAJ acknowledge support from NASA {\jwst} Interdisciplinary
Scientist grants NAG5-12460, NNX14AN10G and 80NSSC18K0200 from GSFC. Work by CJC acknowledges support from the European Research Council (ERC) Advanced Investigator Grant EPOCHS (788113). BLF thanks the Berkeley Center for Theoretical Physics for their hospitality during the writing of this paper. MAM acknowledges the support of a National Research Council of Canada Plaskett Fellowship, and the Australian Research Council Centre of Excellence for All Sky Astrophysics in 3 Dimensions (ASTRO 3D), through project number CE17010001. CNAW acknowledges funding from the {\jwst}/NIRCam contract NASS-0215 to the University of Arizona.
CC is supported by the National Natural Science Foundation of China, No. 11803044, 12173045.


%

\vspace{5mm}
\facilities{
James Webb Space Telescope; Mikulski Archive
}


\software{astropy \citep{AstropyCollaboration2013, AstropyCollaboration2018},  
          Source Extractor \citep{Bertin1996},
          TOPCAT \citep{Taylor2005}
          }




\bibliography{refs}{}
\bibliographystyle{aasjournal}



\end{document}